# Analysis of water injection heat recovery potential of abandoned oil wells to geothermal wells in northern Shaanxi

**Yu Huagui[1,2], Liu Shi[3], Pang Yanyan[3], Wang Peng[4], Gao Qian[1,2]**

( 1. College of New Energy, Xi'an Shiyou University, Xi'an City, Shaanxi Province, 710065, China

2. Engineering Research Center of Smart Energy and Carbon Neutral in Oil & Gas Field, Universities of Shaanxi Province, Xi'an City, Shaanxi Province, 710065, China

3. College of Petroleum Engineering, Xi 'an Petroleum University, Xi 'an 710065, Shaanxi

4. Anton Oilfield Services (Group) Ltd, Beijing, 100102, China)

**Abstract :** The Chang 2 bottom water reservoir area in the western part of northern Shaanxi is one of the core oil-producing areas in the Ordos Basin.One of the main reservoirs is the Chang 2 reservoir of the Triassic Yanchang Formation, which has good physical conditions, active edge and bottom water, and high geothermal gradient. In this paper, the reservoir numerical simulation software CMG is used to simulate the water intake and heat recovery in the target study area, and the heat recovery rate and heat recovery of the three water production methods of direct water production, four injection and one production and one injection and four production under different injection pressures are analyzed. The results show that it is difficult to realize the direct water extraction from the bottom water reservoir. The annual heat recovery of single well of four injection and one production and one injection and four production is converted to the standard coal production between 190 ~ 420 t, so the Chang 2 reservoir in the western part of northern Shaanxi has the potential of water injection and heat recovery.

## 1. Introduction

China 's proven geothermal resources rank second in the world. The reserves of geothermal resources represented by hydrothermal type are about 1,250 billion tons, but the current development level is only 2 %. First of all, from the perspective of resource development, if we vigorously develop geothermal resource projects, the annual exploitation of geothermal resources equivalent to coal equivalent to half of China 's coal consumption in 2015, so the development potential of geothermal resources is beyond doubt. Secondly, from the perspective of the treatment of abandoned oil wells, the number of abandoned oil wells is increasing with the rapid development of China 's oil industry. Abandoned oil wells will emit a large amount of gas and liquid that pollute the environment, which requires a lot of manpower and material resources to improve the strict sealing of wells. The use of structurally complete abandoned oil wells to develop geothermal energy for building heating or refrigeration, industrial production, power generation, etc., can not only achieve energy utilization, but also reduce the drilling cost of new geothermal wells and the cost of sealing abandoned oil wells. In their research, Barbier 's drilling cost of new geothermal wells accounted for 50 % of the total cost of geothermal projects[1]. Therefore, the development direction of abandoned oil wells to geothermal wells is worthy of further study.

In the process of injection and heat recovery, Zhang Jie et al.believed that when using abandoned oil wells to extract heat[2], we can pay attention to the well pattern distribution of oil fields. Choosing reasonable injection-production ratio and well pattern distribution can improve geothermal development. They numerically simulated the heat production methods of the two systems of one injection and one production and two injection and one production. The simulation results show that the heat production performance of the two injection and one production system is better than that of the one injection and one production system under the same system conditions. Guo Tao used the data of Liubei buried hill reservoir in Huabei Oilfield to establish a thermal-water-mechanical multi-field coupling mathematical model[3], simulated the geothermal development of abandoned high-temperature oilfields, and analyzed the sensitivity of geothermal development effect to well pattern, fracturing conditions and working media. The results show that the heat transfer area of one injection four production well network is the largest, and the hydraulic fracturing model leads to an earlier thermal breakthrough, but generates more heat than the non-fracturing model. In this paper, combined with the injection-production heat transfer effect of different well patterns in the research status and the existing well pattern structure in the western part of northern Shaanxi, the STARS module of reservoir numerical simulation software CMG is used to analyze the production potential of direct water production, four injection and one production and one injection and four production in the Chang 2 bottom water reservoir in the western part of northern Shaanxi.2.1

**2. Heat transfer analysis**

*2.1 Geothermal geological conditions in the study area*

*( 1 ) Thermal reservoir temperature conditions*

The terrestrial heat flow value is the heat flow value per unit area of the earth 's surface per unit time. This parameter eliminates the influence of time and is currently the most representative parameter of the earth 's surface temperature. The terrestrial heat flow and temperature gradient can be used to estimate the temperature conditions of thermal reservoirs in a certain range of underground. The more active the tectonic activity is, the higher the terrestrial heat flow value is, and the more stable the structure is, the smaller the terrestrial heat flow value is. The average value of heat flow in the Ordos Basin in central China is 55 ~ 80 mW / m², which is smaller than that in Tibet, but larger than that in Xinjiang and other regions. In addition, the geothermal gradient in various regions of China is 1.5 ~ 4.0 °C / 100m, with an average of 3.2 °C / 100m.In general, the geothermal gradient in the north is larger than that in the south, and the east is larger than the west. The average geothermal gradient in the south is about 2.45 °C / 100m, and the average geothermal gradient in the north is about 3 °C / 100m.

*( 2 ) Regional geological resource reserves*

Geothermal resources in China are mainly derived from hydrothermal geothermal resources and dry hot rocks. The study area is located in the Ordos Basin. The resources in the Ordos Basin have the characteristics of good reservoir conditions, many reservoirs, large thickness and wide distribution.

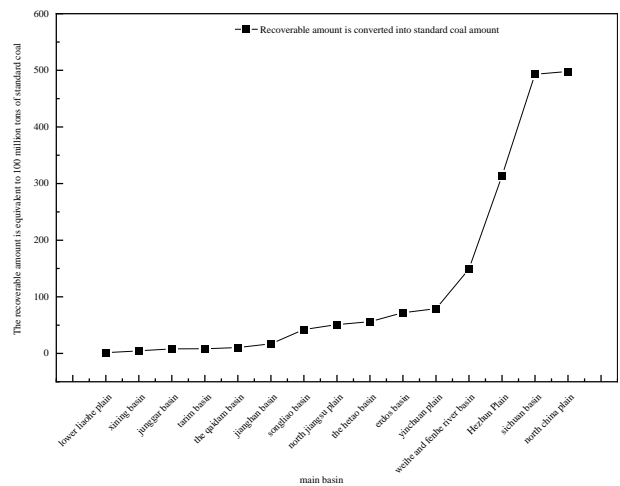

Fig.1 Geothermal fluid recoverable heat reserves of main basins in China

As shown in Figure 1, from the understanding of geothermal resource reserves by Wang Guiling and others[4], the recoverable heat of geothermal fluid in the Ordos Basin is not the highest in the national basins, but the total amount still reaches 308 million tons / year, so it still has great geothermal development potential. In addition, Kang Jing et al.evaluated the geothermal resources in the Yanchang oil and gas area in the southeast of Ordos[5], and the Chang 2 bottom water reservoir in the west of northern Shaanxi also belongs to the geothermal resource enrichment area. In the research status, we mentioned that the area where the reservoir is located can multiply the geothermal mining effect. Among them, Chang 2 is a medium-porosity and low-permeability reservoir with active bottom water edge water, so there is a condition for water extraction and heat extraction after abandonment.

*2.2 Direct water intake from bottom water reservoir*

The water cut of oil and gas fields is high when the water cut is 70-98 %. However, due to the continuous injection and production, the formation and reservoir of oil and gas fields have changed greatly. The current development focus has been developed from a large range of connected remaining oil to a smaller range. When the aquifer exceeds 98 %, these wells will be shut down, and some will even be permanently shut down and become abandoned wells. After the oil and gas resources in the oil and gas wells are exhausted, if there are hydrothermal geothermal resources, the oil layer can be blocked and converted into a water layer, so that the geothermal water is transported to the heating station through the geothermal pipe network, and then the heat exchange with the heat exchanger is carried out to transfer the heat of the geothermal water to the heating circulating water. The heated heating circulating water is then transported to the user through the heating pipe network for use by the user. After the reservoir is fully water-flooded, some will form a bottom water reservoir. Bottom water reservoir refers to the reservoir with large and sheet water at the bottom of the reservoir. This kind of reservoir stops oil production due to abandonment after water injection. However, due to the rich water resources in such reservoirs, it is possible to try to directly take water for heat recovery or inject heat, and change abandoned oil wells into geothermal wells for heat recovery operations.

The northern Shaanxi oil region is a ' three low ' reservoir. Based on the basic data of the Chang 2 reservoir in a target block, the situation of water production and heat extraction is simulated. The reservoir physical parameters are shown in Table 2-1, and the formation heat recovery model is established on this basis.

**Table 1 Reservoir physical parameters table**

| physical parameters | parameter value |
|---|---|
| Porosity | 18% |
| Permeability | $20 \times 10^{-3} \, \mu m^2$ |
| Reservoir top depth | 1200 m |
| Reservoir thickness | 6 m |
| Reservoir bottom depth | 1206 m |
| Reservoir temperature | 45 °C |
| Reservoir pressure | 7 MPa |
| Rock compression coefficient | $0.27 \times 10^{-4} \, MPa^{-1}$ |
| Formation water salinity | 28870 mg/L |

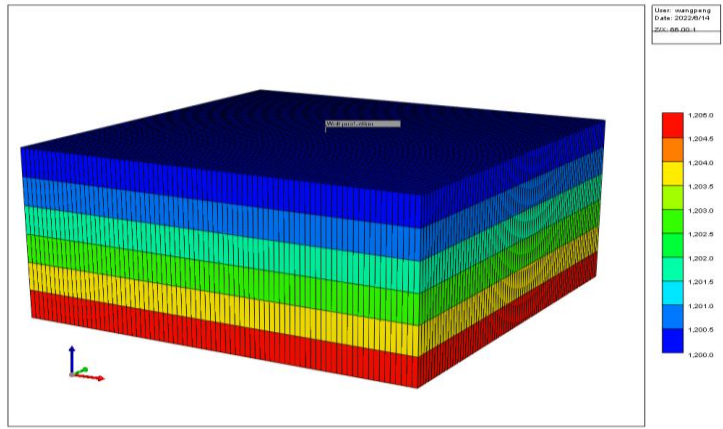

Fig.2 Three-dimensional model of direct water extraction

As shown in Fig.2, the water layer is perforated, and then the water layer is extracted for heat extraction. During water production, the bottom hole flow pressure is controlled to be 0.2 MPa, and each layer is ejected.

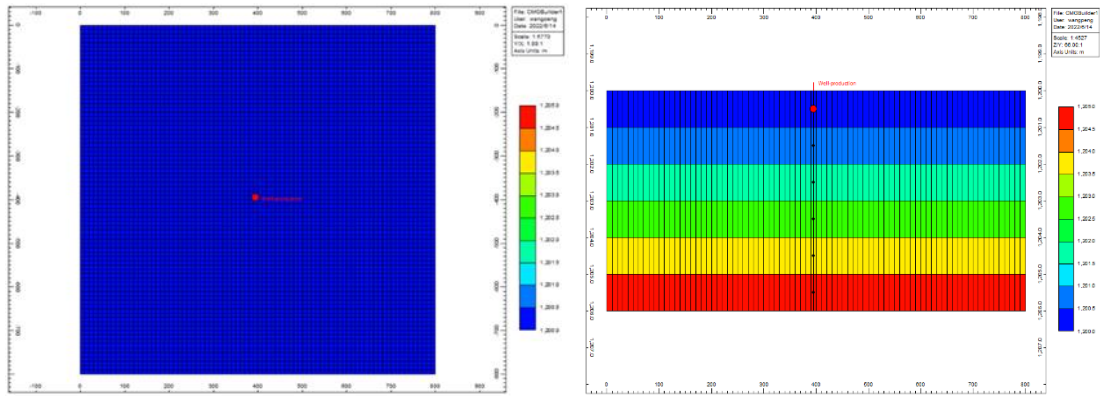

( a ) Perforation water top surface　　　　　　( b ) Perforation water profile

Fig. 3 The top surface and profile of the perforation water heat extraction model

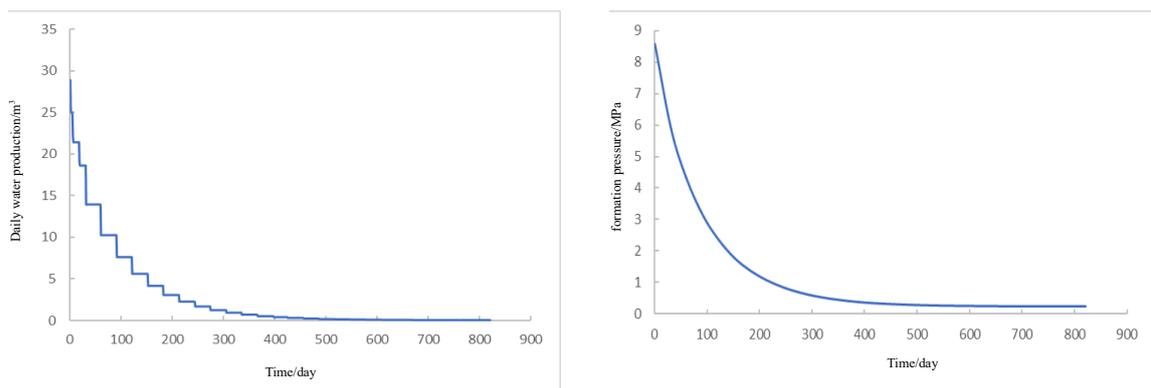

(a) Perforated water intake daily water production　　(b) Perforation water formation pressure change

Fig.4 Changes of daily water production and formation pressure during perforation water heating

The water production and heat extraction simulation is carried out by using the model in Fig.3.Fig.4 shows the change of water production and pressure in the formation without water injection. It can be seen from the simulated production data that the water production decreases rapidly after well opening, from the maximum 28 m$^3$ / d in the initial stage to less than 10 m$^3$ / d in less than 100 days, and decreases to about 1 m$^3$ / d in 11 months. This production curve is very similar to the scenario of depletion development of oilfield in this area. The reason is mainly related to the low permeability physical properties of the reservoir. Therefore, in this low permeability reservoir, relying on natural pressure to extract water and heat will lead to a serious decline in the average pressure of the formation, and it is unsustainable to exploit groundwater only by natural energy. There is a starting pressure gradient in low permeability reservoirs, and there is often a low-speed, high-resistance and non-flowing area between oil and water wells, which is the main reason affecting the development effect of low permeability oilfields.

*2.3 Injection takes heat*

In the analysis of well pattern distribution of Chang 2 reservoir by Liu Xue et al.[6], the well pattern distribution of Chang 2 reservoir in the target area is a more flexible diamond inverted nine-point well pattern. Therefore, the existing abandoned well pattern can meet a variety of injection and production methods. However, it can be seen from the research status that the heat exchange of two injection and one production is greater than that of one injection and one production, and the heat exchange area of one injection and four production well pattern is the largest. Therefore, two well pattern structures of four injection and one production and one injection and four production are selected. Combined with different injection pressures, the heat recovery of Chang 2 bottom water reservoir in western Shaanxi is analyzed. With reference to the above literature, the well spacing of the reservoir is set to 200 m, the reservoir thickness is 50 m, and 2-9 layers are shot.

*2.3.1 Four note one take heat*

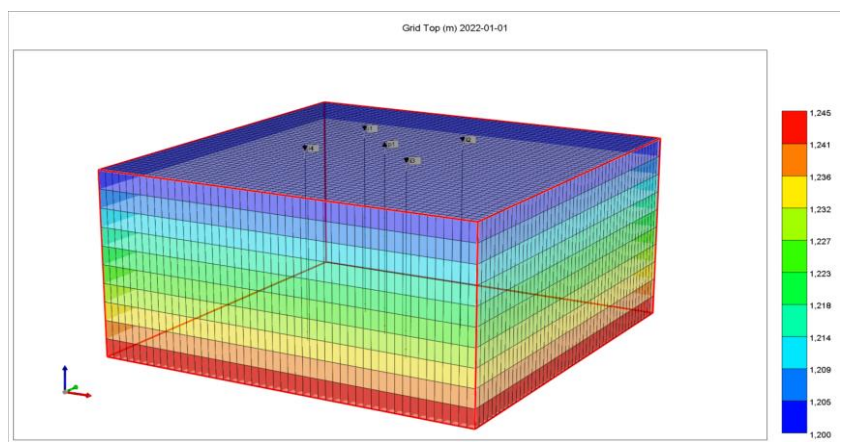

Fig. 5 Three-dimensional model of four-injection and one-mining

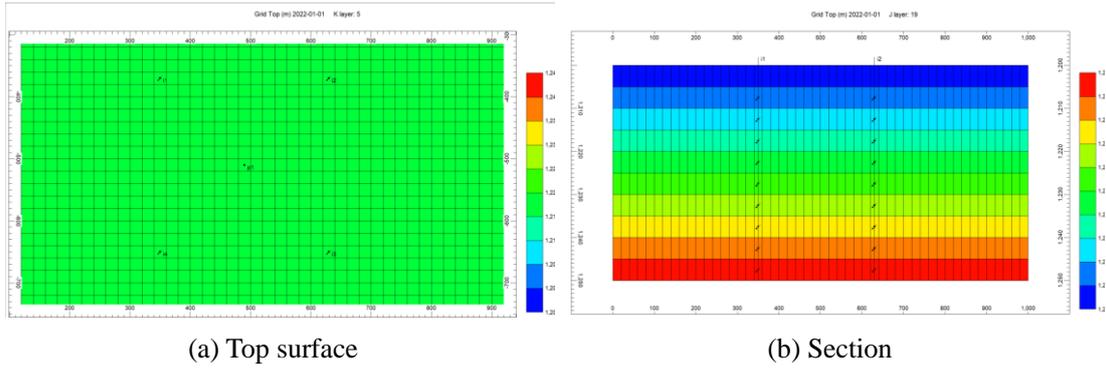

(a) Top surface　　　　　　　　　　　(b) Section

图 6　Four injection one take thermal model top surface, profile

As shown in Fig.5 to Fig.6, p1 is the production well, i1 ~ i4 is the injection well, and the three-dimensional temperature distribution can be seen.

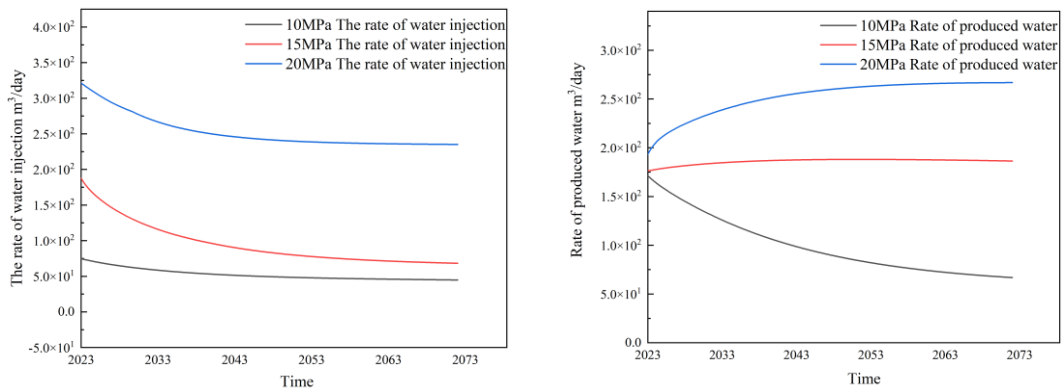

( a ) Daily water injection rate　　　　　　　( b ) Daily water extraction rate

Fig. 7 Changes of daily water production and daily water injection in four-injection and one-mining scheme

It can be seen that the water injection volume has been declining in Fig.7.It should be that as the total injection volume increases, the formation pressure increases and the water injection rate slows down. In addition, it can be seen that the greater the injection pressure, the greater the daily water production rate. When the injection pressure is 15 MPa, the formation pressure can be replenished in time. When the injection pressure is 10 MPa, the injection pressure is too small to replenish the formation pressure in time, so the water production volume decreases first and then remains unchanged. When the injection pressure is 20 MPa, the injection volume is large, the formation is replenished or even increased, and a high permeability channel may be formed[7]. In the case of sufficient formation pressure, the water production rate continues to rise, but limited by the low porosity and low permeability of the reservoir, the water production rate gradually tends to be gentle.

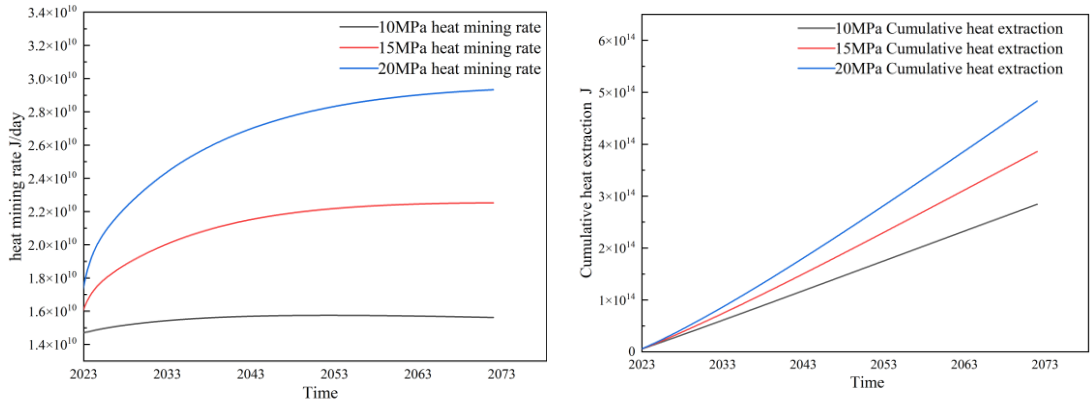

( a ) Daily heat extraction rate　　　　( b ) Cumulative heat extraction

Fig. 8 Four-injection-one-mining scheme Four-injection-one-mining heat recovery rate and heat recovery

It can be seen from Fig.8 that the daily heat recovery rate increases first and then tends to be gentle, which is related to the water recovery rate. The cumulative heat recovery is related to the heat recovery rate and the number of years[8]. The larger the heat recovery rate is, the longer the heat recovery period is, and the greater the cumulative heat recovery is.

*2.3.2 One note four take heat*

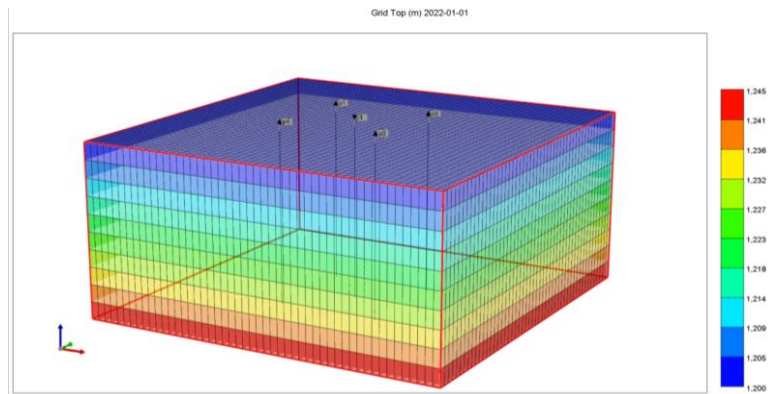

Fig.9 Three-dimensional model of one injection four mining scheme

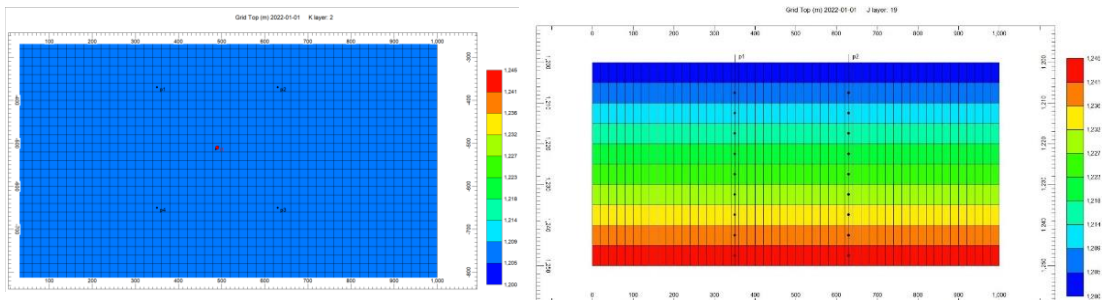

( a ) Top surface　　　　　　　　　　( b ) Section

Fig.10 The top surface and profile of the heat extraction model of one injection four production scheme

P1 ~ P4 is the production well, and i1 is the water injection well. From Fig.9 to Fig.10, the three-dimensional temperature distribution can be seen.

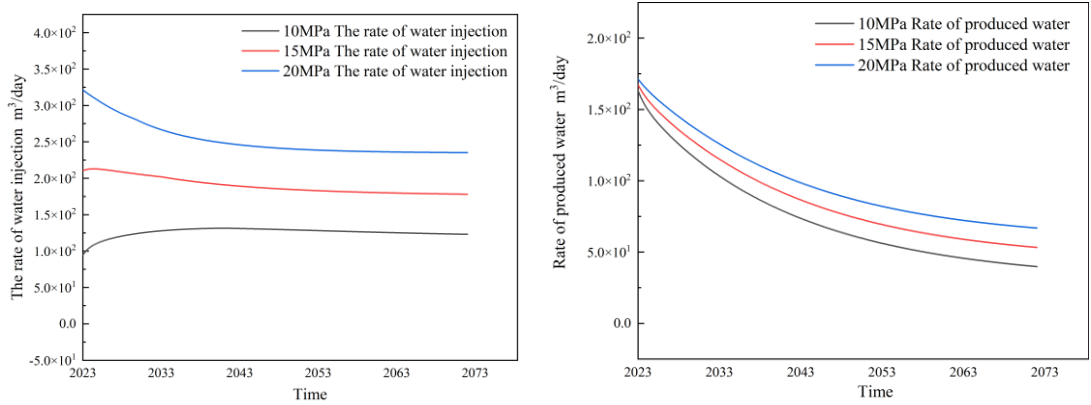

( a ) The rate of water injection　　　　　　( a ) Rate of produced water

Fig. 11 Changes of daily water production and daily water injection in one injection and four mining schemes

It can be seen from Fig.11 that the daily water injection rate of 20 MPa and 15 MPa decreased, but the 15 MPa decreased slowly. This is because after a period of injection, the formation pressure around the injection well gradually increased[9], the pressure around the production well became smaller, the production pressure difference became smaller, and the injection rate became slower. In addition, it can be seen in ( b ) that the greater the pressure, the greater the daily water extraction rate.

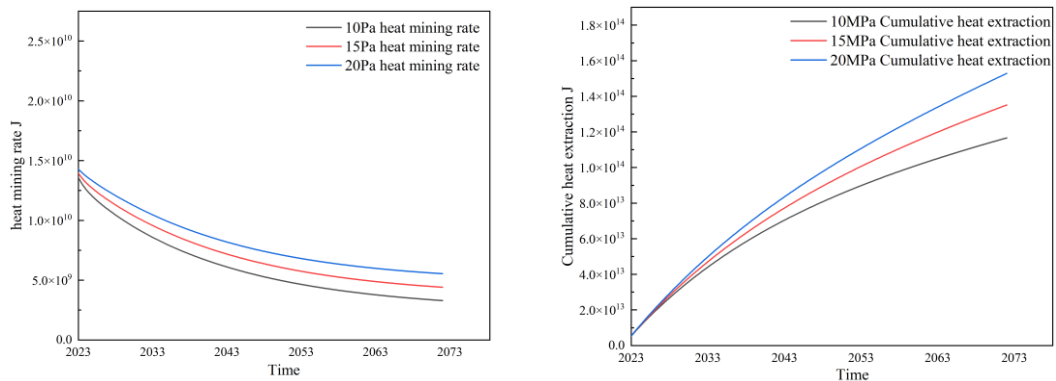

( a ) Daily heat extraction rate　　　　　　( b ) Cumulative heat extraction

Fig.12 Heat recovery rate and heat recovery of one injection four production scheme

From the heat extraction situation in Fig.12, the water extraction rate decreases, the heat extraction rate decreases, and the cumulative heat extraction increases more and more slowly.

The statistics of heat recovery under different injection-production methods and injection pressures are made, as shown in table 2.

Table 2 Different heat recovery models and heat recovery conditions of injection pressure

| injection-production method | injection pressure （MPa） | Single well annual heat production（J） | The annual cumulative heat recovery of the system（J） |
| --- | --- | --- | --- |

| | 10 | 5.68×10¹² | 5.68×10¹² |
|---|---|---|---|
| Four injection one mining | 15 | 7.71×10¹² | 7.71×10¹² |
| | 20 | 9.65×10¹² | 9.65×10¹² |
| | 10 | 2.33×10¹² | 9.33×10¹² |
| One injection four mining | 15 | 2.70×10¹² | 1.08×10¹³ |
| | 20 | 3.06×10¹² | 1.22×10¹³ |

It can be seen that the greater the injection pressure, the greater the annual heat production of single well. Under the same injection pressure, the annual heat production of single well with one injection and four production is greater than that of single well with four injection and one production, but there are four production wells with one injection and four production. Therefore, under the same injection pressure, the annual heat production of the system with four injection and one production is less than that of one injection and four production.

Table 3 Comparison of heat recovery between four-injection-one-mining and one-injection-four-mining

| injection-production method | injection pressure (MPa) | Annual heat extraction (J) | Convert standard coal quantity (t) | Reduced C emissions (t) | Reduced $CO_2$ emissions (t) | Profits (ten thousand yuan) |
|---|---|---|---|---|---|---|
| Four injection one mining | 10 | 5.68×10¹² | 193.8 | 129.8 | 476.1 | 38.8 |
| | 15 | 7.71×10¹² | 263.0 | 176.2 | 646.2 | 52.6 |
| | 20 | 9.65×10¹² | 329.2 | 220.6 | 808.8 | 65.8 |
| One injection four mining | 10 | 9.33×10¹² | 318.2 | 213.2 | 781.7 | 63.6 |
| | 15 | 1.08×10¹³ | 368.6 | 247.0 | 905.5 | 73.7 |
| | 20 | 1.22×10¹³ | 417.0 | 279.4 | 1024.3 | 83.4 |

Using the " reference conversion coefficient of various energy sources and standard coal " issued by Guizhou Provincial Energy Bureau in November 2018, the heat obtained each year is converted into standard coal, which is converted into data such as carbon emissions and carbon dioxide emissions. It can be seen from Table 3 that the amount of standard coal converted from annual heat production has certain economic value, and reduces the carbon emissions corresponding to standard coal, and accelerates the energy development purpose of ' carbon neutralization and carbon peak '.Therefore[10], the abandoned oil wells in the western part of northern Shaanxi have the potential and significance of heat injection.

**3 conclusion**

( 1 ) The edge and bottom water of Chang 2 bottom water reservoir in the western part of northern Shaanxi is more active, and the geothermal temperature condition is better, which has the condition of replacing abandoned wells with geothermal wells for water and heat recovery.

（2）If the direct water extraction of the reformed well is equivalent to depletion mining, the formation pressure drops seriously and cannot be continuously mined.

（3）CMG simulation shows that when the water injection pressure is between 10 ~ 20 MPa, the water injection pressure increases and the average annual heat recovery increases.

（4）In the CMG model, the simulation time of 50 years is longer. In the early stage, there are four production wells with one injection and four production wells, so the total heat exchange is greater than that of four injection and one production wells. However, with the development, the water production rate of one injection and four production wells decreases rapidly, while the water production rate of four injection and one production wells is relatively stable. Therefore, four injection and one production wells have the possibility of long-term development in well reform. However, in the simulated 50 years, the annual average heat production proves that one injection and four production wells are feasible in a short period of time. Injection-production plan.

（5）The average annual heat production of the other two injection-production methods is converted into standard coal production between 190 ~ 420 t. Therefore, whether one injection four production or four injection one production, it is proved that the Chang 2 bottom water reservoir in the western part of northern Shaanxi has the ability to take water and heat.

Author Profile :Yu Huagui, Associate Professor of Xi 'an Petroleum University, mainly engaged in geothermal energy development, CCUS, etc. Email: yuhuagui@xsyu.edu.cn.